\documentclass{article}
\usepackage{epsfig}
\usepackage{color}
\usepackage{amsmath}
\usepackage{amssymb}
\usepackage{lscape}

\makeatletter
\@addtoreset{equation}{section}

\makeatother

\begin{document}

\begin{flushright}
May 2014
\end{flushright}

\begin{center}

\vspace{3cm}

{\LARGE 
\begin{center}
Supersymmetry Breaking and Planar Free Energy

in 

Chern-Simons-matter Theories
\end{center}
}

\vspace{2cm}

Takao Suyama \footnote{e-mail address : suyama@ewha.ac.kr}

\vspace{1cm}

{\it 
Institute for the Early Universe, Ewha Womans University

\vspace{2mm}

Seoul 120-750 KOREA 
}

\vspace{2cm}

{\bf Abstract} 

\end{center}

We investigate a relation between zeros of the partition function and supersymmetry breaking, 
conjectured by Morita and Niarchos, for a family of Chern-Simons-matter theories. 
We analyze the analytic structure of the free energy in the large $N$ limit derived from the resolvent of 
the corresponding matrix model. 
We find that a branch point exists at the value of the 't Hooft coupling 
above which supersymmetry is known to be broken, confirming the conjecture. 

\newpage

\vspace{1cm}

\section{Introduction}

\vspace{5mm}

It has been known since \cite{Witten:1999ds} 
that ${\cal N}=1$ supersymmetric Chern-Simons theory exhibits supersymmetry breaking for a certain 
range of the parameters. 
For example, if the gauge group is ${\rm SU}(N)$ and the Chern-Simons level is $k>0$, then the supersymmetry is broken if and only if 
\begin{equation}
N\ >\ 2k. 
\end{equation}
Strictly speaking, this is the result for ${\cal N}=1$ Yang-Mills-Chern-Simons theory and, since the Yang-Mills term is irrelevant 
in the IR limit, this should be also the case for pure Chern-Simons theory. 
This result was shown in \cite{Witten:1999ds} by calculating the Witten index (in fact, even the number of vacua was calculated). 
The calculation was extended to ${\cal N}=2,3$ theories in \cite{Ohta:1999iv}. 
For these theories, supersymmetry is broken if and only if 
\begin{equation}
N\ >\ k. 
   \label{relation}
\end{equation}

The supersymmetric (Yang-Mills-)Chern-Simons theory can be realized on a system of branes. 
${\cal N}=2$ pure Chern-Simons theory is realized on D3-branes suspended between 
NS5-branes and D5-branes  \cite{Kitao:1998mf}\cite{Bergman:1999na}. 
In this setup, the condition for supersymmetry breaking is examined by the s-rule \cite{Hanany:1996ie} of the brane system. 
The s-rule claims that only a single D3-brane can be suspended between an NS5-brane and a D5-brane without breaking supersymmetry. 
Therefore, the rank of the gauge group is restricted by the number of D5-branes, resulting in the condition (\ref{relation}). 

Curiously, the result for pure Chern-Simons theory had not been extended to more general 
Chern-Simons-matter theories until the study of ABJ theory \cite{Aharony:2008gk} was started. 
ABJ theory is ${\cal N}=6$ Chern-Simons-matter theory with the gauge group ${\rm U}(M)\times{\rm U}(N)$ and the level $(k,-k)$ 
coupled to bi-fundamental matters. 
It was conjectured in \cite{Aharony:2008gk} that the supersymmetry is preserved if 
\begin{equation}
|M-N|\ \le\ k, 
\end{equation}
provided $k>0$, and otherwise the theory is inconsistent, not merely non-supersymmetric. 

The condition of supersymmetry breaking for more general ${\cal N}=2$ Chern-Simons-matter theories was obtained recently in 
\cite{Intriligator:2013lca}. 
For a theory with the gauge group $G$ and the level $k>0$ coupled to matters in the representations $R_i$ of $G$, 
the supersymmetry is broken if and only if 
\begin{equation}
h(G) \ > \ k+\frac12\sum_{i}T_2(R_i),  
   \label{IS}
\end{equation}
where $h(G)$ is the dual Coxeter number of $G$ and $T_2(R)$ is the quadratic index of $R$. 
Note that $h(G)=N$ for $G={\rm SU}(N)$. 

Recently, it was conjectured in \cite{Morita:2011cs} that there would be 
an interesting connection between the supersymmetry breaking discussed above and the value of the partition function. 
The claim is that the supersymmetry is broken if and only if the partition function vanishes. 
Supporting arguments were given in \cite{Morita:2011cs}. 
For pure Chern-Simons theory \cite{Morita:2011cs} and ABJ theory 
\cite{Awata:2012jb}, this conjecture was shown to be correct by the explicit calculation of the partition 
function. 
Note that the vanishing partition function and the inconsistency of ABJ theory is reminiscent of the ${\rm SU}(2)$ anomaly 
\cite{Witten:1982fp}. 

\vspace{5mm}

In this paper, we examine the above conjecture for a family of Chern-Simons-matter theories. 
We consider ${\cal N}=3$ Chern-Simons theories with the gauge group ${\rm U}(N)$ and the level $k$ coupled to $N_f$ fundamental 
matters and $n$ adjoint matters. 
For theories including them, an integral formula for the partition function was obtained in \cite{Kapustin:2009kz}. 
The evaluation of the partition function can be done in a suitable large $N$ limit using matrix model techniques. 
Since the planar resolvent for those theories was obtained in \cite{Suyama:2011yz}\cite{Suyama:2012uu}, 
their free energy can be analyzed. 
Then, the conjecture of \cite{Morita:2011cs}, 
translated into a relation between supersymmetry breaking and the position of a singularity 
of the free energy, can be examined. 
Note that the conditions above for the supersymmetry breaking make sense in a suitable large $N$ limit. 

We will show in this paper that the free energy has a branch point at the place which is expected from the condition (\ref{IS}). 
The analysis in this paper can be extended to a more general family of Chern-Simons-matter theories, based on the results 
obtained in \cite{Suyama:2013fua}. 

This paper is organized as follows. 
In section \ref{review}, we review known results on the conjecture based on the explicit evaluation of the partition function. 
A large $N$ version of the conjecture is also stated. 
In section \ref{planar general}, 
we investigate the analytic structure of the free energy as a function of the 't Hooft coupling $\lambda$ using the planar 
resolvent. 
According to whether the supersymmetry is broken or preserved, the free energy is shown to have a branch point at the right position, 
or to be holomorphic for $\lambda>0$. 
Section \ref{discuss} is devoted to discussion. 
A puzzle for the case $n=\frac12$, corresponding to ${\cal N}=3$ Chern-Simons theory coupled to a single (anti-)symmetric matter, 
is also mentioned. 
Appendices contain some details.

\vspace{1cm}

\section{Known results}  \label{review}

\vspace{5mm}

We begin with a brief review on 
known results on pure Chern-Simons theory, Chern-Simons theory coupled to fundamental matters, 
and ABJ theory for all of which the partition function was analyzed exactly. 
The results are compatible with the conjecture of \cite{Morita:2011cs}. 
It will be also shown that the conjectured relation can be investigated in terms of the planar free energy in the large $N$ limit. 

\vspace{5mm}

\subsection{Pure Chern-Simons theory}

\vspace{5mm}

Consider ${\cal N}=2$ pure Chern-Simons theory with the gauge group ${\rm U}(N)$ and the level $k$. 
The partition function is \cite{Kapustin:2009kz} 
\begin{equation}
Z(N,k)\ =\ \prod_{m=1}^N\left( \sin\frac{\pi m}{k} \right)^{N-m}, 
   \label{pureCS-Z}
\end{equation}
up to an overall factor which is not important in the following. 
Apparently, the product vanishes for $N>k$ while it is non-zero otherwise. 
This pattern coincides with the pattern of the supersymmetry breaking reviewed in the introduction. 
Therefore, for the case of pure Chern-Simons theory, $Z(N,k)$ vanishes if and only if the supersymmetry is broken. 

\vspace{5mm}

It is interesting to notice that the free energy $F(N,k):=-\log Z(N,k)$ has a well-define `t Hooft limit \cite{Morita:2011cs} 
\begin{eqnarray}
F(\lambda) 
&:=& -\lim_{N\to\infty}\frac1{N^2}\log Z(N,k) \nonumber \\
&=& -\int_0^1dx\,(1-x)\log\sin(\pi\lambda x), 
\end{eqnarray}
where $\lambda:=\frac Nk$ is fixed. 
Due to zeros of $Z(N,k)$ at $N>k$, $F(\lambda)$ is expected to have a singular behavior at $\lambda=1$. 
The value $F(1)$ is, however, finite\footnote{
The free energy in \cite{Morita:2011cs} is different from the one in this paper by a constant. 
}: 
\begin{equation}
F(1)\ =\ \frac12\log2. 
\end{equation}
Instead, $\lambda=1$ is a branch point of $F(\lambda)$. 
This may become more transparent by rewriting $F(\lambda)$ as 
\begin{equation}
F(\lambda)\ =\ -\frac12\log\sin(\pi\lambda)+\frac1{2\lambda^2}\int_0^\lambda dx\,(2\lambda-x)\pi x\cot(\pi x). 
\end{equation}
The first term has branch points at $\lambda\in\mathbb{Z}$. 
The second term also has branch points at $\lambda\in\mathbb{Z}\backslash\{0\}$ due to poles of $\cot(\pi x)$. 
The discontinuity of $F(\lambda)$ on an interval $1<\lambda<2$ is 
\begin{eqnarray}
F(\lambda+i0)-F(\lambda-i0) 
&=& \pi i-\frac1{2\lambda^2}\oint_1 dx\,(2\lambda-x)\pi x\cot(\pi x) \nonumber \\ 
&=& \pi i\left( 1-\frac1\lambda \right)^2. 
   \label{disc}
\end{eqnarray}

Note that $F(\lambda)$ is finite for $\lambda\notin\mathbb{Z}$. 
This suggests that an expected sign of the supersymmetry breaking in the planar free energy should be the presence of a branch point, 
not necessarily a divergent behavior. 

\vspace{5mm}

\subsection{Chern-Simons theory with fundamental matters}

\vspace{5mm}

Pure Chern-Simons theory may couple to a number of matters. 
Consider a Chern-Simons-matter theory with the gauge group ${\rm U}(N)$ with the level $k>0$. 
The maximum amount of supersymmetry allowed for the theory is ${\cal N}=3$ \cite{Gaiotto:2008sd}. 
In this case, the matters form ${\cal N}=4$ hypermultiplets. 
In this subsection, we consider hypermultiplets in the fundamental representation of ${\rm U}(N)$. 
Each hypermultiplet consists of one ${\cal N}=2$ fundamental multiplet and one ${\cal N}=2$ anti-fundamental multiplet. 
Let $N_f$ be the number of the fundamental hypermultiplets, and $Z(N,k,N_f)$ be the partition function of the theory. 
It was shown in \cite{Kapustin:2010mh} that (schematically) 
\begin{equation}
Z(N,k,N_f)\ \propto\ Z(N-1,k,N_f-1). 
\end{equation}
This relation, combined with the result on pure Chern-Simons theory, implies that $Z(N,k,N_f)=0$ if and only if 
\begin{equation}
N\ >\ k+N_f. 
\end{equation}
Indeed, this is the condition for the supersymmetry breaking for this theory 
derived from the s-rule of a brane system. 
Appendix \ref{brane} reviews a brane realization of the ${\cal N}=3$ theory. 

\vspace{5mm}

\subsection{ABJ theory}

\vspace{5mm}

The partition function of ABJ theory was obtained as a finite-dimensional integral via localization \cite{Kapustin:2009kz}. 
The integral was then performed explicitly in \cite{Awata:2012jb}. 
Let the gauge group be ${\rm U}(N)\times{\rm U}(N+M)$ and the levels be $k>0$ and $-k$, respectively. 
The partition function turns out to have an overall factor 
\begin{equation}
\prod_{j=1}^{M-1}(q)_j\ \propto\ \prod_{j=1}^{M-1}\left( \sin\frac{\pi j}{k} \right)^{M-j}, 
\end{equation}
which has the same functional form as (\ref{pureCS-Z}) for pure Chern-Simons theory. 
Therefore, the partition function vanishes if and only if $M>k$, which is the same condition for the inconsistency of ABJ theory reviewed 
in the introduction. 

\vspace{5mm}

The planar free energy of ABJM theory \cite{Aharony:2008ug}, corresponding to $M=0$, was determined in \cite{Drukker:2010nc}. 
The third derivative of the planar free energy $F(\lambda)$ is given as 
\begin{equation}
\partial_\lambda^3F(\lambda)\ =\ -\frac{128\pi^6}{\kappa(\kappa^2+16)}\frac1{K(\frac{i\kappa}4)}, 
\end{equation}
where $\kappa$ is related to $\lambda$ via 
\begin{equation}
\frac{d\lambda}{d\kappa}\ =\ \frac1{4\pi^2}K\left( \mbox{$\frac{i\kappa}4$} \right), 
   \label{ABJM}
\end{equation}
and $K(k)$ is the complete elliptic integral of the first kind which can be written as 
\begin{equation}
K(k)\ =\ \int_0^{\frac\pi2}\frac{d\theta}{\sqrt{1-k^2\sin^2\theta}}\ =\ \frac\pi2F\left( \mbox{$\frac12,\frac12$},1;k^2 \right). 
\end{equation}
The right-hand side of (\ref{ABJM}) is positive for real $\kappa$. 
Therefore, any positive real $\lambda$ corresponds to a positive real $\kappa$. 
Since $F(\lambda(\kappa))$ is a holomorphic function of $\kappa$ on the positive real $\kappa$ axis, 
there is no branch point of $F(\lambda)$ for any positive real $\lambda$. 
This is compatible with the conjecture of \cite{Morita:2011cs} since no supersymmetry breaking is expected for ABJM theory.

\vspace{1cm}

\section{Planar free energy and resolvent} \label{planar general}

\vspace{5mm}

In this section, we discuss the planar free energy of ${\cal N}=3$ Chern-Simons-adjoint theories 
\cite{Suyama:2012uu} which may couple to fundamental 
matters. 
The gauge group is ${\rm U}(N)$ and the level is $k$. 
Let $n$ be the number of adjoint hypermultiplets, and $N_f$ be the number of fundamental hypermultiplets. 
The theory is defined on $S^3$. 
The partition function can be given in terms of a finite-dimensional integral
\begin{equation}
Z(N,k,n,N_f)\ =\ \int \prod_{i=1}^Ndu_i\,e^{-S[u]}, 
   \label{partition fn}
\end{equation}
where 
\begin{equation}
S[u]\ =\ \sum_{i=1}^N\frac{k}{4\pi i}u_i^2-\sum_{i<j}\log\left[ \sinh^2\frac{u_i-u_j}2 \right]
 +n\sum_{i\ne j}\log\left[ \cosh\frac{u_i-u_j}2 \right]+N_f\sum_{i=1}^N\log\left[ \cosh\frac{u_i}2 \right], 
\end{equation}
via the localization \cite{Kapustin:2009kz}. 
In the limit 
\begin{equation}
N,k,N_f\ \to\ \infty, \hspace{5mm} \lambda\ :=\ \frac Nk \ \mbox{ and }\  n_f\ :=\ \frac{N_f}k \ \mbox{ fixed,}
\end{equation}
the integral (\ref{partition fn}) is dominated by $\{\bar{u}_i\}$ satisfying the equations $\partial_{u_i}S[u]=0$, that is, 
\begin{equation}
\frac k{2\pi i}u_i+\frac{N_f}2\tanh\frac{u_i}2\ =\ \sum_{j\ne i}\coth\frac{u_i-u_j}2-n\sum_{i}\tanh\frac{u_i-u_j}2. 
   \label{saddle}
\end{equation}
If $k$ is purely imaginary, it is natural to expect that the $\{\bar{u}_i\}$ would lie on the real axis, forming a continuous 
segment in the large $N$ limit. 
By the symmetry of the equations, the segment should be $[-\alpha,\alpha]$ for some $\alpha>0$. 
The distribution of $\{\bar{u}_i\}$ for general $k$ would be given by a complex $\alpha$ determined by the analytic continuation of $k$. 

It is convenient to introduce a new variable $z:=e^u$. 
In terms of $z$, the equations (\ref{saddle}) become 
\begin{equation}
\frac k{2\pi i}\log z_i+\frac{N_f}2\frac{z_i-1}{z_i+1}\ =\ \sum_{j\ne i}\frac{z_i+z_j}{z_i-z_j}-n\sum_i\frac{z_i-z_j}{z_i+z_j}. 
   \label{saddle2}
\end{equation}

The resolvent $v(z)$ defined as 
\begin{equation}
v(z)\ :=\ \lim_{N\to\infty}\frac tN\sum_{j}\frac{z+z_j}{z-z_j}, \hspace{5mm} t\ :=\ 2\pi i\lambda
\end{equation}
is assumed to define an analytic function on $\mathbb{C}\backslash[a,b]$ where $a:=e^{-\alpha},b:=e^{\alpha}$. 
The line segment $[a,b]$ is a branch cut of $v(z)$. 
The 't Hooft coupling $t$ is given by $v(z)$ as 
\begin{equation}
t\ =\ v(\infty)\ =\ -v(0). 
   \label{tHooft}
\end{equation}
In terms of $v(z)$, the equations (\ref{saddle2}) can be written as 
\begin{equation}
2\log y+2\pi in_f\frac{y-1}{y+1}\ =\ v(y^+)+v(y^-)-2n\,v(-y), 
   \label{saddle3}
\end{equation}
where 
\begin{equation}
y^\pm\ :=\ y\pm i0, \hspace{5mm} y\in[a,b]. 
\end{equation}

Let $v_0(z)$ be the solution of (\ref{saddle3}) with $n_f=0$. 
Then, $v(z)$ is given as 
\begin{equation}
v(z)\ =\ v_f(z)+v_0(z), 
\end{equation}
where $v_f(z)$ has a branch cut at the same position as the one of $v_0(z)$, and satisfies 
\begin{equation}
2\pi in_f\frac{y-1}{y+1}\ =\ v_f(y^+)+v_f(y^-)-2n\,v_f(-y).  
\end{equation}
It was shown in \cite{Suyama:2012uu} that $v_0(z)$ can be given in the form 
\begin{equation}
v_0(z)\ =\ \int_{-\infty}^0d\xi\,v_0(z,\xi), 
\end{equation}
where $v_0(z,\xi)$ satisfies 
\begin{equation}
-\frac2{\xi-1}\frac{y-1}{y-\xi}\ =\ v_0(y^+,\xi)+v(y^-,\xi)-2n\,v(-y,\xi). 
   \label{saddle-general}
\end{equation}
It is interesting to notice that $v_0(z,\xi)$ also gives $v_f(z)$ as 
\begin{equation}
v_f(z)\ =\ 2\pi in_fv_0(z,-1). 
\end{equation}
This fact was used recently in \cite{Grassi:2014vwa}. 

\vspace{5mm}

The planar free energy is given as 
\begin{equation}
F_{n,n_f}(\lambda)\ :=\ -\lim_{N\to\infty}\frac1{N^2}S[\bar{u}]. 
   \label{planar F}
\end{equation}
It can be shown that the first derivative of $\lambda^2F_{n,n_f}(\lambda)$ is given in terms of the resolvent $v(z)$ as 
\begin{equation}
\partial_\lambda( \lambda^2 F_{n,n_f}(\lambda))
 \ =\ \int_{z_0}^a\frac{dz}{2\pi i}\frac1z\left[ -v(z)+n\,v(-z)+\log z+\pi in_f\frac{z-1}{z+1} \right]
 +\frac{\delta f}{\delta\varphi(u_0)}\Big|_{\varphi_\lambda}. 
   \label{integral rep}
\end{equation}
Note that the last term will be irrelevant in the following. 
For the details, see Appendix \ref{integral-rep_app}. 

We are interested in the analytic structure of $F_{n,n_f}(\lambda)$. 
The above integral representation (\ref{integral rep}) gives $F_{n,n_f}(\lambda)$ as a function of $a$, one of the branch points of 
$v(z)$. 
Since (\ref{tHooft}) relates $\lambda$ and $a$, we have a parametric representation of $F_{n,n_f}(\lambda)$. 
The integral representation (\ref{integral rep}) indicates that $F_{n,n_f}(\lambda)$, as a function of $a$, 
may have branch points at $a=0,-1$. 
However, it will be shown in the following that these points do not correspond to real finite $\lambda$. 
This implies that a possible branch point of $F_{n,n_f}(\lambda)$ at a real finite $\lambda$ 
should come from a branch point of $a(\lambda)$ given by (\ref{tHooft}). 

\vspace{5mm}

In the following, we will examine the analytic structure of $a(\lambda)$, or equivalently $a(t)$, for various $n$ and $n_f$. 
If a branch point exists on the real axis in the $\lambda$-plane, or on the imaginary axis in the $t$-plane, the position will be 
compared with the one expected from the pattern of supersymmetry breaking discussed in section \ref{review}, or 
with the condition (\ref{IS}).

\vspace{5mm}

\subsection{Pure Chern-Simons theory}

\vspace{5mm}

The simplest case $n,n_f=0$ corresponds to pure Chern-Simons theory. 
This theory was examined in section \ref{review} using the exact partition function. 
In this subsection, we reproduce the same results based on the planar resolvent. 
The results obtained here suggest how to investigate more complicated theories. 

\vspace{5mm}

\begin{figure}
\begin{center}
\includegraphics{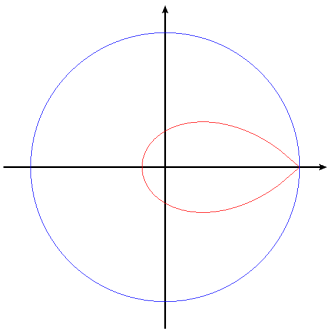}
\caption{The curve (red) in $a$-plane to which the imaginary axis of $t$-plane is mapped.
Since $|a|\le1$ is assumed, the curve lies inside the unit circle. }
\end{center}
\label{fig1}
\end{figure}

The solution of (\ref{saddle-general}) is \cite{Suyama:2012uu} 
\begin{equation}
v_0(z,\xi)\ =\ -\frac1{\xi-1}\frac{z-1}{z-\xi}+\frac1{z-\xi}\frac{\sqrt{(z-a)(z-b)}}{\sqrt{(\xi-a)(\xi-b)}}. 
   \label{resolvent_pureCS}
\end{equation}
Note that $v_0(z,\xi)$ does not have a pole at $z=\xi$ due to $ab=1$, as it should be the case. 

The 't Hooft coupling $t$ is given as 
\begin{equation}
t(a)\ =\ -\int_{-\infty}^0\frac{d\xi}{\xi}\left[ \frac1{\sqrt{(\xi-a)(\xi-b)}}+\frac1{1-\xi} \right]. 
\end{equation}
In fact, this integration can be done explicitly. 
As a result, one obtains the following relation 
\begin{equation}
2e^{\frac12t}\ =\ \sqrt{a}+\frac1{\sqrt{a}}. 
\end{equation}
This defines a map from $t$-plane to $a$-plane. 
The imaginary axis of the $t$-plane, corresponding to the physical values, is mapped to a curve in the $a$-plane depicted in Figure 1. 
Note that, by definition, $a$ is restricted to $|a|\le1$. 
Since the curve does not pass through the origin, there is no branch point on the curve. 

A subtlety appears around $a=1$. 
To see this, recall $a=e^{-\alpha}$, and write $t$ in terms of small $\alpha$. 
One obtains 
\begin{equation}
t\ =\ \frac14\alpha^2+O(\alpha^4). 
\end{equation}
This indicates that $a(t)$ has a branch point at $t=0$. 
Indeed, such a branch point was found in section \ref{review}. 

Figure 1 shows that, by increasing $t$ in the imaginary direction, the curve goes around the origin, and at $t=2\pi i$ 
it comes back to $a=1$. 
Therefore, there is another branch point of the same kind at $t=2\pi i$. 
Here, the relation between $t$ and $\alpha$ is 
\begin{equation}
t-2\pi i\ = \ \frac14( \alpha-2\pi i )^2+O((\alpha-2\pi i)^4). 
\end{equation}

\begin{figure}
\begin{center}
\includegraphics{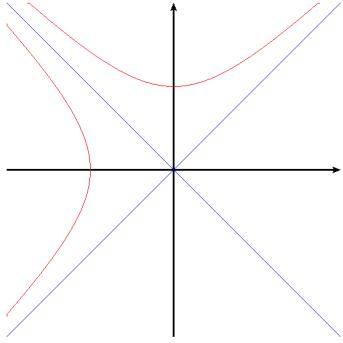}
\label{pureCS2}
\caption{The curves in $\alpha$-plane near $\alpha=0$. 
The left red curve corresponds to the case ${\rm Re}(t)>0$, while the upper red curve corresponds to the case ${\rm Re}(t)<0$. }
\end{center}
\end{figure}

Suppose that $t$ has a small real part corresponding to a small imaginary part of $\lambda$. 
Depending on the sign of ${\rm Re}(t)$, the curve in the $t$-plane passing near $t=2\pi i$ is mapped into different 
curves in the $\alpha$-plane, as 
depicted in Figure 2. 
For a small negative ${\rm Re}(t)$ (a small positive ${\rm Re}(\lambda)$), ${\rm Re}(\alpha)$ becomes positive, implying that 
$|a|$ becomes greater than 1. 
This means that the role of $a$ and $b$ should be exchanged. 

Due to the fact that the image of the map from the $t$-plane to the $a$-plane strongly depends on the sign of ${\rm Re}(t)$, 
$F_{0,0}(\lambda)$ has a discontinuity. 
The integral representation (\ref{integral rep}) implies that the discontinuity for $1<\lambda<2$ is 
\begin{eqnarray}
{\rm Disc}\,\partial_\lambda(\lambda^2F_{0,0}(\lambda))
&=& \frac12\oint_C\frac{dz}{2\pi i}\frac{v_0(z)}{z} \nonumber \\
&=& 2\pi i(\lambda-1). 
\end{eqnarray}
Integrating this with respect to $\lambda$, and using the fact that there is no discontinuity at $\lambda=1$, one can reproduce 
(\ref{disc}). 

\vspace{5mm}

We have found that the branch point on the real $\lambda$ axis closest to the origin is $\lambda=1$, and a branch cut appears for 
$\lambda>1$. 
This is equivalent to 
\begin{equation}
N\ >\ k, 
\end{equation}
which coincides with the condition for supersymmetry breaking.

\vspace{5mm}

\subsection{Chern-Simons theory coupled to fundamental multiplets}

\vspace{5mm}

Next, consider the case $n=0$ and $n_f>0$. 
As mentioned at the beginning of this section, the resolvent $v(z)$ has the form 
\begin{equation}
v(z)\ =\ v_0(z)+v_f(z). 
\end{equation}
Here $v_0(z)$ is the same as the resolvent of pure Chern-Simons theory discussed above. 
$v_f(z)$ is obtained from (\ref{resolvent_pureCS}) as \cite{Suyama:2010hr} 
\begin{equation}
v_f(z)\ = \ 2\pi in_f\left[ \frac12\frac{z-1}{z+1}-\frac1{z+1}\frac{\sqrt{(z-a)(z-b)}}{\sqrt{(1+a)(1+b)}} \right]. 
\end{equation}
The 't Hooft coupling $t$ for this theory is given in terms of $a$ as 
\begin{equation}
t\ =\ w+\pi in_f\left[ 1-e^{-\frac12w} \right], \hspace{5mm} 2e^{\frac12w}\ =\ \sqrt{a}+\frac1{\sqrt{a}}. 
\end{equation}
As in pure Chern-Simons theory, $a(w)$ has a branch point at $w=2\pi i$. 
It turns out that this point lies on a curve 
\begin{equation}
w_{\rm R}e^{\frac12w_{\rm R}}\ =\ \pi n_f\sin\frac{w_I}2, 
\end{equation}
where $w=w_{\rm R}+iw_{\rm I}$, in the $w$-plane to which the imaginary axis of the $t$-plane is mapped. 
See Figure 3. 
One can check that $w=2\pi i$ corresponds to $t=2\pi i(1+n_f)$, and no other branch point exists between this and the origin. 
If ${\rm Im}(t)$ increases further, $F_{0,n_f}(\lambda)$ exhibits a discontinuity, as in pure Chern-Simons theory. 
The condition for the presence of a branch cut 
\begin{equation}
\lambda\ >\ 1+n_f
\end{equation}
can be written as 
\begin{equation}
N\ >\ k+N_f, 
\end{equation}
which coincides with the known condition for supersymmetry breaking. 

\begin{figure}
\begin{center}
\includegraphics{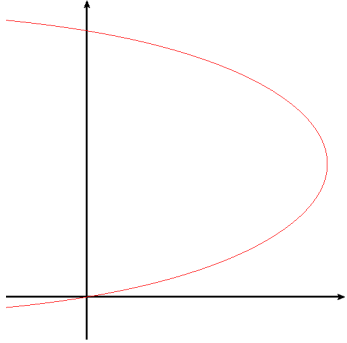}
\label{fCS}
\end{center}
\caption{The curve for the case $n_f=1$ in the $w$-plane corresponding to the imaginary axis of the $t$-plane. 
It intersects with the imaginary axis at $w=0,2\pi i$. }
\end{figure}

\vspace{5mm}

\subsection{Chern-Simons-adjoint theory: $n>1$} \label{non-deg}

\vspace{5mm}

We consider the case $n>1$ and $n_f=0$. 
The theory with $n=1$ requires a different treatment, so it will be discussed later. 
The planar resolvent $v(z)$ was obtained in \cite{Suyama:2012uu}. 
The functional form of $v(z)$ is complicated, but the relation between $t$ and $a$ derived via (\ref{tHooft}) is rather simple: 
\begin{equation}
t\ =\ \frac1{n-1}\int_{-\infty}^0\frac{d\xi}{\xi}\left[ \frac i{2e\sin\frac{\pi\nu}2}G(\xi)+\frac1{1-\xi} \right]. 
   \label{CSAd-t}
\end{equation}
Here $\nu$ is related to $n$ as 
\begin{equation}
n\ =\ -\cos\pi\nu. 
\end{equation}
$e$ is a constant depending on $a$ and $n$, and $G(z)$ is a function given in terms of theta functions. 
For the details, see Appendix \ref{details}. 

\begin{figure}
\begin{center}
\includegraphics{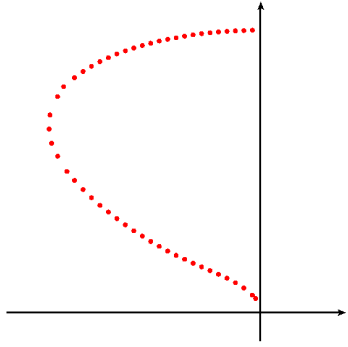}
\end{center}
\label{CSAd1}
\caption{Points in the $\tau$-plane at which ${\rm Re}(t)=0$ are plotted for the case $\nu=1-0.2i$. 
The curve ${\rm Re}(t)=0$ starts from the origin and intersects with the imaginary axis at $\tau=10i$. }
\end{figure}

The integral (\ref{CSAd-t}) defines $t$ as a function of $a$. 
It turns out to be convenient to change the parameter from $a$ to $\tau$ defined as 
\begin{equation}
a\ =\ \frac{\vartheta_2^0(\tau)}{\vartheta_3^0(\tau)}. 
   \label{tau}
\end{equation}
The behavior of $t(\tau)$ can be investigated numerically. 
Several points in the $\tau$-plane at which ${\rm Re}(t)$ vanishes are plotted in Figure 4. 
The line connecting these points starts from the origin, and approaches to a point on the imaginary axis of the $\tau$-plane at which $t$ 
diverges. 
This divergence is due to the factor $e^{-1}$ in (\ref{CSAd-t}). 
Since $n>1$, $\nu$ must be chosen as 
\begin{equation}
\nu\ =\ 1-2i\mu, \hspace{5mm} \mu\ >\ 0. 
\end{equation}
Then, $e$ is given as 
\begin{equation}
e\ =\ -\frac{\vartheta_1(i\mu\tau,\tau)}{\vartheta_0(i\mu\tau,\tau)}. 
\end{equation}
Therefore, $t$ diverges when $\vartheta_1(i\mu\tau,\tau)=0$. 
For example, if 
\begin{equation}
\tau\ \in\ i\mu^{-1}\mathbb{Z}, 
\end{equation}
then $t$ diverges. 
In fact, there are other values of $\tau$ at which $t$ diverges, but those which are 
not on the imaginary axis of $\tau$-plane does not correspond 
to purely imaginary $t$. 

One can check that there is no branch point on the curve corresponding to $\lambda>0$. 
This suggests that the supersymmetry is not broken for ${\cal N}=3$ Chern-Simons-adjoint theory with $n>1$. 
Interestingly, this is the same conclusion from 
the condition (\ref{IS}) of supersymmetry breaking for ${\cal N}=2$ Chern-Simons theory coupled to $2n$ adjoint 
multiplets. 
We expect that the condition of supersymmetry breaking for an ${\cal N}=3$ theory is the same as the one 
for the corresponding ${\cal N}=2$ theory which shares the same gauge group and the same matter content. 
In fact, an ${\cal N}=3$ theory is obtained from an ${\cal N}=2$ theory by taking a suitable limit of parameters. 
Therefore, if the supersymmetry is preserved for generic values of the parameters for the ${\cal N}=2$ theory, then even in the limit 
the supersymmetry should be preserved. 
Note that the ${\cal N}=3$ theory is not at the boundary of the parameter space of ${\cal N}=2$ theories, so there would not be a 
possibility that the supersymmetric vacuum would disappear after taking the limit. 
Indeed, we have seen that for pure Chern-Simons theory and Chern-Simons theory with flavors, the pattern of supersymmetry breaking is 
the the same between ${\cal N}=3$ theory and ${\cal N}=2$ theory. 

\vspace{5mm}

Note that the above numerical study of $t(a)$ confirms the expectation that $\lambda\to\infty$ corresponds to $a\to a_*$ 
mentioned in \cite{Suyama:2012uu}. 
As a result, the expectation value of the Wilson loop is bounded by a constant even in the large $\lambda$ limit, which is a quite 
different behavior from the Wilson loop in ABJM theory.

\vspace{5mm}

\subsection{Chern-Simons-adjoint theory: $n=1$}

\vspace{5mm}

\begin{figure}
\begin{center}
\includegraphics{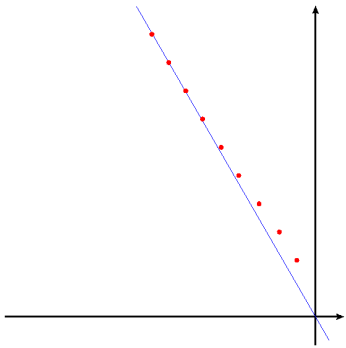}
\label{degenerate}
\caption{Some points with ${\rm Re}(t)=0$ are plotted on the $\tau$-plane. 
The blue line corresponds to $\arg(\tau)=\frac23\pi$ to which the curve ${\rm Re}(t)=0$ approaches asymptotically. }
\end{center}
\end{figure}

The equation (\ref{saddle3}) for this theory was solved in \cite{Suyama:2011yz}. 
In fact, it turned out that, instead of $v(z)$, a combination $v(z)-v(-z)$ may have a rather simple expression as follows: 
\begin{equation}
v(z)-v(-z)\ =\ 4z\int_a^b\frac{dx}{2\pi}\frac{\log x}{z^2-x^2}\frac{\sqrt{(z^2-a^2)(z^2-b^2)}}{\sqrt{|(x^2-a^2)(x^2-b^2)|}}. 
\end{equation}
A disadvantage of using this combination is that the 't Hooft coupling $t$ cannot be obtained by simply setting $z$ in the above 
expression to be $0$ or $\infty$. 
Instead, $t$ can be recovered as 
\begin{equation}
t\ =\ \frac12\oint_C\frac{dz}{2\pi i}\frac{v(z)-v(-z)}{z}, 
   \label{n=1}
\end{equation}
where the contour $C$ encircles the branch cut $[a,b]$ of $v(z)$. 
The integral can be evaluated numerically. 
It is again convenient to introduce $\tau$ defined as (\ref{tau}). 
Several points in the $\tau$-plane at which ${\rm Re}(t)=0$ are plotted in Figure 5. 
This shows that the line to which the imaginary $t$ axis is mapped starts from the origin and goes to infinity, not like in the case 
$n>1$. 
Indeed, the integral (\ref{n=1}) can be estimated when ${\rm Im}(\tau)$ is large and positive. 
The leading behavior is 
\begin{equation}
t\ \sim\ \frac{\pi i}{96}\tau^3. 
\end{equation}
This implies that the line ${\rm Re}(t)=0$ corresponds, in the $\tau$-plane, to a line which asymptotically approaches 
$\arg(\tau)=\frac{2\pi}3$, as the plot also indicates. 
In terms of $\alpha=-\log a$, this can be written as 
\begin{equation}
t\ \sim\ \frac2{3\pi^2}\alpha^3, 
\end{equation}
which was also obtained in \cite{Suyama:2011yz} for real $\alpha$. 

It can be checked that there is no branch point for all $\lambda>0$. 
Assuming again that the condition (\ref{IS}) can be used for ${\cal N}=3$ theories, 
this result is compatible with the conclusion of (\ref{IS}) that there is no supersymmetry breaking 
for this theory.

\vspace{5mm}

\subsection{Chern-Simons-matter theories with fundamental matters}

\vspace{5mm}

Finally, we consider the case $n,n_f>0$. 
For the case $n>1$, $v_f(z)$ is determined from $v_0(z,\xi)$. 
The additional contribution to the 't Hooft coupling $t$ from $v_f(z)$ is 
\begin{equation}
-\frac{2\pi in_f}{n-1}\left[ \frac i{2e\sin\frac{\pi\nu}2}G(-1)+\frac12 \right]. 
   \label{from flavors}
\end{equation}
However, for the case $n=1$, it is not so straightforward to determine such a contribution since $v_0(z,\xi)$ was not determined. 
Fortunately, it was shown in \cite{Grassi:2014vwa} that the expression (\ref{from flavors}) has a well-defined limit $n\to1$. 
Therefore, we can also use (\ref{from flavors}) for the investigation of the case $n=1$. 

It can be checked numerically that there is no branch point for these theories. 
This result is compatible with the conclusion of (\ref{IS}) that there is no supersymmetry breaking in these theories.

\vspace{1cm}

\section{Discussion} \label{discuss}

\vspace{5mm}

We have investigated the analytic structure of the planar free energy of a family of Chern-Simons-matter theories. 
The planar free energy is a function of the 't Hooft coupling $\lambda$ which 
can be obtained from the planar resolvent of the corresponding matrix model. 
As deduced from the conjecture \cite{Morita:2011cs}, the free energy turned out to have 
a branch point at the value of $\lambda$ above which 
the supersymmetry is broken. 

The relation between the vanishing of the partition function and the supersymmetry breaking seems to be natural from the point of view 
of \cite{Suyama:2012kr}. 
In \cite{Suyama:2012kr}, 
Chern-Simons-matter theories are regarded as a low energy effective theory of a Yang-Mills theory coupled to matters 
including fundamental matters with "axial masses." 
When these masses are heavy, they are integrated out, leaving an appropriate Chern-Simons term. 
Instead, if those masses are non-zero but small, it is possible to argue that there exists a non-perturbative superpotential, 
as discussed in \cite{Aharony:1997bx}, and the supersymmetry is broken due to a non-zero vacuum energy. 
It was found in \cite{Suyama:2012kr} 
that the condition for the existence of non-perturbative superpotential and the condition for the supersymmetry breaking 
for the corresponding Chern-Simons-matter theory are closely related. 
In the limit of infinite masses, where the theory is reduced to a Chern-Simons-matter theory, the vacuum energy induced by the 
superpotential seems to become infinite, and therefore, the partition function would naturally vanish. 

To the extent discussed so far, the conjecture of \cite{Morita:2011cs} turned out to be correct. 
However, there seems to be a puzzle if one analyzes the free energy of ${\cal N}=3$ Chern-Simons theory with the gauge group ${\rm U}(N)$ 
coupled to a single hypermultiplet in the (anti-)symmetric representation of ${\rm U}(N)$. 
The planar resolvent of this theory is a solution of (\ref{saddle3}) with $n=\frac12$ and $n_f=0$ \cite{Suyama:2013fua}. 
If, as discussed in subsection \ref{non-deg}, 
the condition (\ref{IS}) is valid for this theory, it implies that the supersymmetry is broken if and only if 
\begin{equation}
N\ >\ 2k, \hspace{5mm} \mbox{or} \hspace{5mm} \lambda\ >\ 2. 
\end{equation}
However, the investigation of (\ref{CSAd-t}) for the case $n=\frac12$ does not show any sign of a branch point at $\lambda=2$. 
This might not be a counterexample of the conjecture of \cite{Morita:2011cs} 
since it is the planar free energy that is under consideration. 
Recall that, for pure Chern-Simons theory, the free energy for a finite $N$ is divergent when $N>k$, but the planar free energy is 
finite for most of that range of $\lambda$. 
This is simply because $\log x$ is integrable around $x=0$. 
For pure Chern-Simons theory, the branch cut still remains after the large $N$ limit, indicating the non-analyticity. 
However, it might be possible that, for Chern-Simons theory with an (anti-)symmetric matter, branch cuts from various logarithmic 
singularities would cancel among them in the large $N$ limit. 
Actually, it was observed that there is a sign of a branch cut at $\lambda=3$. 
It is interesting to clarify this issue. 
It is also interesting to generalize the analysis in this paper to more general family of Chern-Simons-matter theories based on the 
results in \cite{Suyama:2013fua}. 

\vspace{2cm}

{\bf \Large Acknowledgments}

\vspace{5mm}

This work was supported in part by the Research fund no. 1-2010-2469-001 by Ewha Womans University. 

\vspace{2cm}

{\bf \Large Appendix}

\appendix

\vspace{1cm}

\section{Brane realization ${\cal N}=3$ Chern-Simons theory with fundamental matters} \label{brane}

\vspace{5mm}

Recall the realization of ${\cal N}=3$ pure Chern-Simons theory in Type IIB string theory \cite{Kitao:1998mf}\cite{Bergman:1999na}. 
There are two NS5-branes parallel to each other, and $N$ D3-branes are suspended between the NS5-branes. 
Introduce $k$ D5-branes intersecting with one of the NS5-branes. 
These branes extend in the following manner: 
\begin{center}
\begin{tabular}{c|ccc|ccc|c|ccc}
 & 0 & 1 & 2 & 3 & 4 & 5 & 6 & 7 & 8 & 9 \\ \hline 
NS5 & $\bigcirc$ & $\bigcirc$ & $\bigcirc$ & $\bigcirc$ & $\bigcirc$ & $\bigcirc$ & & & &  \\
D3 & $\bigcirc$ & $\bigcirc$ & & & & & $\bigcirc$ & & & \\ 
D5 & $\bigcirc$ & $\bigcirc$ & $\bigcirc$ & $\bigcirc$ & $\bigcirc$ & & & & & $\bigcirc$ 
\end{tabular}
\end{center}
By the web deformation \cite{Bergman:1999na}, the intersecting NS5-brane and D5-branes become a $(1,k)$5-brane, and 
then one realizes ${\cal N}=2$ pure Chern-Simons theory on the D3-branes. 
By suitably rotating the $(1,k)$5-brane, the supersymmetry is enhanced to ${\cal N}=3$ \cite{Kitao:1998mf}\cite{Bergman:1999na}. 

This ${\cal N}=3$ theory may couple to fundamental hypermultiplets by introducing another set of D5-branes, denoted as D5'-branes. 
To see that these D5'-branes do not break supersymmetry, let us return to the brane system with two NS5-branes and $N$ D3-branes 
only. 
If $N_f$ D5'-branes are introduced in the following manner, 
\begin{center}
\begin{tabular}{c|ccc|ccc|c|ccc}
 & 0 & 1 & 2 & 3 & 4 & 5 & 6 & 7 & 8 & 9 \\ \hline 
NS5 & $\bigcirc$ & $\bigcirc$ & $\bigcirc$ & $\bigcirc$ & $\bigcirc$ & $\bigcirc$ & & & &  \\
D3 & $\bigcirc$ & $\bigcirc$ & & & & & $\bigcirc$ & & & \\ 
D5' & $\bigcirc$ & $\bigcirc$ & $\bigcirc$ & & & & & $\bigcirc$ & $\bigcirc$ & $\bigcirc$ 
\end{tabular}
\end{center}
the D5'-branes does not break ${\cal N}=4$ supersymmetry preserved by the NS5-D3 system \cite{Hanany:1996ie}. 
Since the ${\cal N}=3$ supersymmetry of pure Chern-Simons theory discussed above is a subgroup of this ${\cal N}=4$ supersymmetry, 
after adding $k$ D5-branes one realizes ${\cal N}=3$ Chern-Simons theory coupled to $N_f$ fundamental hypermultiplets. 
By applying the s-rule to this brane system, one obtains the condition for the supersymmetry breaking 
\begin{equation}
N\ >\ k+N_f. 
\end{equation}

\vspace{1cm}

\section{Integral representation of planar free energy} \label{integral-rep_app}

\vspace{5mm}

Define a functional 
\begin{eqnarray}
f[\varphi] 
&:=& \frac1{4\pi i}\int du\,\varphi(u)u^2-\frac12\int du\,\varphi(u)\int dv\,\varphi(v)\log\sinh^2\frac{u-v}2 \nonumber \\
& & +n\int du\,\varphi(u)\int dv\,\varphi(v)\log\cosh\frac{u-v}2+n_f\int du\,\varphi(u)\log\cosh\frac{u}2. 
\end{eqnarray}
Note that a suitable regularization is understood. 
Let $\varphi_\lambda(u)$ be a function and let $\Gamma_\lambda$ be a constant such that they extremize 
\begin{equation}
f_\Gamma[\varphi]\ :=\ f[\varphi]+\Gamma\left[ \int du\,\varphi(u)-\lambda \right]. 
\end{equation}
Note that $\Gamma_\lambda$ is given as 
\begin{equation}
\Gamma_\lambda\ =\ -\frac{\delta f}{\delta\varphi(u)}\Big|_{\varphi_\lambda}, \hspace{5mm} u\in\mbox{supp}(\varphi_\lambda). 
\end{equation}
It turns out that the planar free energy (\ref{planar F}) is given as 
\begin{equation}
F_{n,n_f}(\lambda)\ =\ \lambda^{-2}f_\Gamma[\varphi_\lambda], 
\end{equation}
and the distribution function $\rho(u)$ of $\bar{u}_i$ in section \ref{planar general} is given as 
\begin{equation}
\rho(u)\ =\ \lambda^{-1}\varphi_\lambda(u). 
\end{equation}
We assume that $\mbox{supp}(\rho)$ is a line segment connecting $u=-\alpha$ and $u=\alpha$. 

Due to the relation 
\begin{equation}
\partial_\lambda f_\Gamma[\varphi_\lambda]\ =\ -\Gamma_\lambda, 
\end{equation}
one can show that 
\begin{eqnarray}
\partial_\lambda( \lambda^2F_{n,n_f}(\lambda) )
&=& \frac{\delta f}{\delta\varphi(-\alpha)}\Big|_{\varphi_\lambda} \nonumber \\
&=& \int_{u_0}^{-\alpha}du\,\frac{d}{du}\frac{\delta f}{\delta\varphi(u)}\Big|_{\varphi_\lambda}
    +\frac{\delta f}{\delta\varphi(u_0)}\Big|_{\varphi_\lambda}, 
   \label{eq-app}
\end{eqnarray}
where $u_0$ is some complex number. 
The first term in the last line can be written as 
\begin{eqnarray}
& & \int_{u_0}^{-\alpha}du\left[ -\int dv\,\varphi_\lambda(v)\coth\frac{u-v}2+n\int dv\,\varphi_\lambda(v)\tanh\frac{u-v}2 
    +\frac1{2\pi i}u+\frac{n_f}2\tanh\frac u2 \right] \nonumber \\
&=& \int_{z_0}^a\frac{dz}{2\pi i}\frac1z\left[ -v(z)+n\,v(-z)+\log z+\pi in_f\frac{z-1}{z+1} \right], 
\end{eqnarray}
where $z=e^u$, $a=e^{-\alpha}$, and $v(z)$ is the resolvent defined in section \ref{planar general}. 
The second term in the last line of (\ref{eq-app}) is a function of $a$. 
If $u_0$ is chosen such that it is far from the segment $[a,b]$, it is a holomorphic function of $a$.

\vspace{1cm}

\section{Details on Chern-Simons-adjoint theories} \label{details}

\vspace{5mm}

The function $G(z)$ appeared in (\ref{CSAd-t}) is a holomorphic function on $\mathbb{C}\backslash[a,b]$. 
It has a branch cut on $[a,b]$ where it satisfies 
\begin{equation}
G(y^+)+G(y^-)-2n\,G(-y)\ =\ 0. 
\end{equation}
Under suitable assumptions on the analyticity, $G(z)$ can be determined uniquely. 
In terms of a new coordinate $u$ defined as 
\begin{equation}
z\ =\ a\,{\rm sn}(u,k), \hspace{5mm} k\ =\ \frac ab, 
\end{equation}
$G(z)$ can be given explicitly as 
\begin{equation}
G(z(u))\ =\ -i\left[ e^{\frac12\pi i\nu}G_+(u)-e^{-\frac12\pi i\nu}G_+(-u) \right], 
\end{equation}
where 
\begin{equation}
G_+(u)\ :=\ -ia\frac{(\vartheta^0_3)^2}{\vartheta^0_0\vartheta_0(\frac{\varepsilon}{2K})}
 \frac{\vartheta_0(\frac u{2K})\vartheta_1(\frac{u-\varepsilon}{2K})}
 {\vartheta_2(\frac{u}{2K})\vartheta_3(\frac u{2K})}e^{-\pi i(1-\nu)\frac u{2K}}
\end{equation}
and 
\begin{equation}
\varepsilon\ :=\ i(1-\nu)K'. 
\end{equation}
$K=K(k)$ and $K'=K'(k)$ are the complete elliptic integrals. 

The parameter $e$ in (\ref{CSAd-t}) is defined as 
\begin{equation}
e\ :=\ a\,{\rm sn}(\varepsilon,k). 
\end{equation}
This is the value of the $z$-coordinate at which $G_+(u)$ vanishes.


\vspace{2cm}










\begin{thebibliography}{99}

\bibitem{Witten:1999ds} 
  E.~Witten,
  ``Supersymmetric index of three-dimensional gauge theory,''
  In *Shifman, M.A. (ed.): The many faces of the superworld* 156-184
  [hep-th/9903005].

\bibitem{Ohta:1999iv} 
  K.~Ohta,
  ``Supersymmetric index and s rule for type IIB branes,''
  JHEP {\bf 9910}, 006 (1999)
  [hep-th/9908120].

\bibitem{Kitao:1998mf} 
  T.~Kitao, K.~Ohta and N.~Ohta,
  ``Three-dimensional gauge dynamics from brane configurations with (p,q) - five-brane,''
  Nucl.\ Phys.\ B {\bf 539}, 79 (1999)
  [hep-th/9808111].

\bibitem{Bergman:1999na} 
  O.~Bergman, A.~Hanany, A.~Karch and B.~Kol,
  ``Branes and supersymmetry breaking in three-dimensional gauge theories,''
  JHEP {\bf 9910}, 036 (1999)
  [hep-th/9908075].

\bibitem{Hanany:1996ie}
  A.~Hanany and E.~Witten,
  ``Type IIB superstrings, BPS monopoles, and three-dimensional gauge dynamics,''
  Nucl.\ Phys.\ B {\bf 492} (1997) 152
  [hep-th/9611230].

\bibitem{Aharony:2008gk} 
  O.~Aharony, O.~Bergman and D.~L.~Jafferis,
  ``Fractional M2-branes,''
  JHEP {\bf 0811}, 043 (2008)
  [arXiv:0807.4924 [hep-th]].

\bibitem{Intriligator:2013lca} 
  K.~Intriligator and N.~Seiberg,
  ``Aspects of 3d N=2 Chern-Simons-Matter Theories,''
  JHEP {\bf 1307}, 079 (2013)
  [arXiv:1305.1633 [hep-th]].

\bibitem{Morita:2011cs} 
  T.~Morita and V.~Niarchos,
  ``F-theorem, duality and SUSY breaking in one-adjoint Chern-Simons-Matter theories,''
  Nucl.\ Phys.\ B {\bf 858}, 84 (2012)
  [arXiv:1108.4963 [hep-th]].

\bibitem{Awata:2012jb} 
  H.~Awata, S.~Hirano and M.~Shigemori,
  ``The Partition Function of ABJ Theory,''
  Prog.\  Theor.\  Exp.\  Phys.\ , 053B04 (2013)
  [arXiv:1212.2966].

\bibitem{Witten:1982fp} 
  E.~Witten,
  ``An SU(2) Anomaly,''
  Phys.\ Lett.\ B {\bf 117}, 324 (1982).

\bibitem{Kapustin:2009kz} 
  A.~Kapustin, B.~Willett and I.~Yaakov,
  ``Exact Results for Wilson Loops in Superconformal Chern-Simons Theories with Matter,''
  JHEP {\bf 1003}, 089 (2010)
  [arXiv:0909.4559 [hep-th]].

\bibitem{Suyama:2011yz} 
  T.~Suyama,
  ``Eigenvalue Distributions in Matrix Models for Chern-Simons-matter Theories,''
  Nucl.\ Phys.\ B {\bf 856}, 497 (2012)
  [arXiv:1106.3147 [hep-th]].

\bibitem{Suyama:2012uu} 
  T.~Suyama,
  ``On Large N Solution of N=3 Chern-Simons-adjoint Theories,''
  Nucl.\ Phys.\ B {\bf 867}, 887 (2013)
  [arXiv:1208.2096 [hep-th]].

\bibitem{Suyama:2013fua} 
  T.~Suyama,
  ``A Systematic Study on Matrix Models for Chern-Simons-matter Theories,''
  Nucl.\ Phys.\ B {\bf 874}, 528 (2013)
  [arXiv:1304.7831 [hep-th]].

\bibitem{Gaiotto:2008sd} 
  D.~Gaiotto and E.~Witten,
  ``Janus Configurations, Chern-Simons Couplings, And The theta-Angle in N=4 Super Yang-Mills Theory,''
  JHEP {\bf 1006}, 097 (2010)
  [arXiv:0804.2907 [hep-th]].

\bibitem{Kapustin:2010mh} 
  A.~Kapustin, B.~Willett and I.~Yaakov,
  ``Tests of Seiberg-like Duality in Three Dimensions,''
  arXiv:1012.4021 [hep-th].

\bibitem{Aharony:2008ug} 
  O.~Aharony, O.~Bergman, D.~L.~Jafferis and J.~Maldacena,
  ``N=6 superconformal Chern-Simons-matter theories, M2-branes and their gravity duals,''
  JHEP {\bf 0810}, 091 (2008)
  [arXiv:0806.1218 [hep-th]].

\bibitem{Drukker:2010nc} 
  N.~Drukker, M.~Marino and P.~Putrov,
  ``From weak to strong coupling in ABJM theory,''
  Commun.\ Math.\ Phys.\  {\bf 306}, 511 (2011)
  [arXiv:1007.3837 [hep-th]].

\bibitem{Grassi:2014vwa} 
  A.~Grassi and M.~Marino,
  ``M-theoretic matrix models,''
  arXiv:1403.4276 [hep-th].

\bibitem{Suyama:2010hr} 
  T.~Suyama,
  ``On Large N Solution of Gaiotto-Tomasiello Theory,''
  JHEP {\bf 1010}, 101 (2010)
  [arXiv:1008.3950 [hep-th]].

\bibitem{Suyama:2012kr} 
  T.~Suyama,
  ``Supersymmetry Breaking in Chern-Simons-matter Theories,''
  JHEP {\bf 1207}, 008 (2012)
  [arXiv:1203.2039 [hep-th]].

\bibitem{Aharony:1997bx} 
  O.~Aharony, A.~Hanany, K.~A.~Intriligator, N.~Seiberg and M.~J.~Strassler,
  ``Aspects of N=2 supersymmetric gauge theories in three-dimensions,''
  Nucl.\ Phys.\ B {\bf 499}, 67 (1997)
  [hep-th/9703110].

\end{thebibliography}
\end{document}